\documentclass{aastex}          %% The manuscript based on AASTeX v5.x
\usepackage{spr-astr-addons}    %% mimicing ASTR journal style
\usepackage{natbib}
\bibpunct{(}{)}{;}{a}{}{,}

\newcommand{\sigmalos}{\sigma_{\rm los}}
\newcommand{\sigmapart}{\sigma_{\rm part}}
\newcommand{\sigmaorb}{\sigma_{\rm orb}}
\newcommand{\halfmass}{R_{\rm hm}}

\newcommand{\msun}{~{\rm M}_\odot}
\newcommand{\mdyn}{M_{\rm dyn}}
\newcommand{\mcl}{M_{\rm cl}}
\newcommand{\kms}{km\,s$^{-1}$}
\newcommand{\binfrac}{F_{\rm M}}
\newcommand{\rsun}{~{\rm R}_\odot}

\begin{document}
%% Article title
%
\title{How do binaries affect the derived dynamical mass of a star cluster?}

%% Running heads
\shorttitle{How do binaries affect the dynamical mass measurement of a star cluster?}
\shortauthors{Kouwenhoven \& de Grijs}

%% Author and Affilations
\author{Kouwenhoven, M.B.N.\altaffilmark{1}} 
\and 
\author{de Grijs, R.\altaffilmark{1,2}}
%\affil{}
%\email{} %% non-output

%% Alternate Affilations
\altaffiltext{1}{Department of Physics and Astronomy, University of Sheffield, Hicks Building, Hounsfield Road, Sheffield S3~7RH, United Kingdom}
\altaffiltext{2}{National Astronomical Observatories, Chinese Academy of Sciences, 20A Datun Road, Chaoyang District, Beijing 100012, P.R.~China}
%\altaffiltext{3}{}

% Abstract
\begin{abstract}
The dynamical mass of a star cluster can be derived from the virial theorem, using the measured half-mass radius and line-of-sight velocity dispersion of the cluster. However, this dynamical mass may be a significant overestimation of the cluster mass if the contribution of the binary orbital motion is not taken into account. In these proceedings we describe the mass overestimation as a function of cluster properties and binary population properties, and briefly touch the issue of selection effects. We find that for clusters with a {\em measured} velocity dispersion of $\sigmalos\ga 10$~\kms{} the presence of binaries does not affect the dynamical mass significantly. For clusters with  $\sigmalos\la 1$~\kms{} (i.e., low-density clusters), the contribution of binaries to $\sigmalos$ is significant, and may result in a major dynamical mass overestimation. The presence of binaries may introduce a downward shift of $\Delta \log (L_V/\mdyn) = 0.05-0.4$ in the $\log (L_V/\mdyn)$ vs. age diagram.
\end{abstract}

% Keywords
\keywords{Star clusters: general --- methods: numerical --- binaries: general}

%%  Please use labels (\label, \ref) for section, figure, table, 
%%  equation  reference. Use \cite for bibliography references.
%

%_____________________________________________________________________________

\section{Introduction}%\label{s:?}

An estimate for the mass of star clusters can be obtained from the virial theorem, using the projected half-mass radius $\halfmass$ and line-of-sight velocity dispersion $\sigmalos$. This dynamical mass estimate, $\mdyn$, is obtained using the equation derived by \cite{spitzer1987}:
\begin{equation} \label{equation:spitzer}
  \mdyn = \eta \, \frac{\halfmass  \sigmalos^2}{G} \,,
\end{equation}
where $\eta \approx 9.75$ is a dimensionless proportionality constant. The derivation of $\mdyn$ using the expression above is valid under the following assumptions: (1) the cluster dynamics are described by the Plummer model, (2) all stars are single and of equal mass, (3) the cluster is in virial equilibrium, and (4) no selection effects are present. Dynamical mass estimates for numerous clusters have been obtained this way \citep[e.g.,][]{mandushev1991,smith2001,maraston2004,bastian2006,larsen2007,moll2007}. When binary stars are present, however, Eq.~(\ref{equation:spitzer}) results in an overestimation of the cluster mass. 

Observations have shown that the majority of the field stars are part of a binary or multiple system \citep{duquennoy1991,fischermarcy1992}. It is believed that most (if not all) binary stars are formed in binary systems, which is supported by both observations \citep{mathieu1994,mason1998,kobulnicky2007,kouwenhoven_adonis,kouwenhoven_recovery} and theory \citep{goodwinkroupa2005}. Stars in binary systems exhibit not only motion in the gravitational potential of the cluster ({\em particle motion} $\sigmapart$), but also {\em orbital motion} $\sigmaorb$ in the binary system. Eq.~(\ref{equation:spitzer}) is applicable for $\sigmapart$ (the centre-of-mass motion of the binaries), but results in an overestimation if $\sigmalos$, the superposition of $\sigmapart$ and $\sigmaorb$, is measured. In this paper we therefore address the question: ``How do binaries affect the dynamical mass of a star cluster?''

%_____________________________________________________________________________

\section{Method and terminology}%\label{s:?}

\begin{table}[tb]
  \caption{The default properties of the model used in our analysis, which we refer to as model~R. In our analysis we vary the properties of modelled star clusters in order to find the effect of these changes on the derived dynamical mass $\mdyn$. At the bottom of the table we list the line-of-sight velocity dispersion $\sigmalos$ of the individual stars, $\sigmapart$ of the centre-of-mass motion of the binaries, and $\sigmaorb$ of solely the orbital motion of the binary components. Each value represents the width of the best-fitting Gaussian.
\label{table:twomodels}}
  \begin{tabular}{ll}
    \hline
    \hline
    Property         & Model R \\
    \hline
    Model            & Plummer \\
    Half-mass radius & $\halfmass=5$~pc\\
    Particles $N=S+B$& $N=18\,600$ \\
    Total mass       & $\mcl=10^4\msun$ \\
    Mass segregation & No \\
    Virial equilibrium& Yes \\
    \hline 
    Primary mass     & $f_{{\rm Kroupa}; 0.08-20 \msun}(M_1)$ \\
    Binary fraction  & $\binfrac=100\%$ \\
    Mass ratio       & $f_q(q) = 1$; $0<q\leq 1$ \\
    Eccentricity     & $f_e(e) = 2e$; $0\leq e < 1$ \\
    Orbital size     & $f_{\rm Opik}(a)$; $10\rsun-0.02$~pc \\
    Orientation      & Random \\
    \hline
    $\sigmalos$ (measured)  & 1.20 \kms{}\\
    $\sigmapart$ \mbox{(centres-of-mass)} & 0.91 \kms{}\\
    $\sigmaorb$ (binaries)  & 0.14 \kms{}\\
    \hline
    \hline
  \end{tabular}
\end{table}

We evaluate the effect of binarity on $\mdyn$ using numerical simulations. We use the STARLAB package \citep[e.g.,][]{ecology4} to model star clusters with different structural and binary population properties. The default properties of our cluster model are listed in Table~\ref{table:twomodels}; we refer to these as 'model~R'. We determine the best-fitting velocity dispersion from the velocity distributions, and do not weigh by mass or luminosity.

In the introduction we described three velocity dispersions: the {\em measured} line-of-sight velocity dispersion $\sigmalos$, the particle (centre-of-mass) velocity dispersion $\sigmapart$, and the binary orbital velocity dispersion $\sigmaorb$. Whether binarity is important depends on which of these $\sigmapart$ or $\sigmaorb$ dominate the measured $\sigmalos$. We discriminate between three types of clusters:
\begin{itemize}
\item Particle-dominated clusters. The measured $\sigmalos$ is dominated by $\sigmapart$; Eq.~(\ref{equation:spitzer}) is a reasonable approximation and $\mdyn/\mcl \la 110\%$.
\item Intermediate-type clusters. The contribution of $\sigmapart$ and $\sigmaorb$ to $\sigmalos$ is comparable; the dynamical mass overestimation is $110\% \la \mdyn/\mcl \la 200\%$.
\item Binary-dominated clusters. The measured $\sigmalos$ is dominated by $\sigmaorb$. The dynamical mass overestimation is significant: $\mdyn/\mcl \ga 200\%$.
\end{itemize}
Typically, binary-dominated clusters have $\sigmalos \la 1$ \kms{}, while particle-dominated clusters have $\sigmalos \ga 10$~\kms{} (see Sect.~\ref{section:danger}).

When changing certain properties of the binary population in a star cluster, such as the increasing the average mass ratio $\langle q \rangle$ or binary fraction $\binfrac$, the total mass of the cluster increases. The latter increase results in larger particle velocity, which should not be fully attributed to the change in the distribution mass ratio distribution $f(q)$ or in $\binfrac$. The increase in particle velocity is partially caused by the increased cluster mass. The latter affect can be compensated for by adjusting the number of particles $N=S+B$ in the cluster. In this way the total cluster mass ``before'' and ``after'' the change in the binary population is equal. The effect of the changed binary property can now be studied accurately. Throughout these proceedings we keep the total mass for each comparison fixed to $10^4 \msun$, a typical value for open cluster-like objects.

%_____________________________________________________________________________

\section{Dependence on cluster properties}%\label{s:?}

In the sections below we describe how the derived dynamical mass depends on the structural parameters of a cluster and on its stellar mass distribution.

%_____________________________________________________________________________

\subsection{Mass distribution and number of particles $N$}

Eq.~(\ref{equation:spitzer}) assumes that all stars are single, equal-mass stars. In reality, the stellar mass spectrum is defined by the mass function, such as the \cite{kroupa2001} IMF. The effect of the mass distribution on the dynamical mass determination, in particular as a function of time, is studied in detail by \cite{fleck2006}, and will not be discussed further here.

The number of particles $N=S+B$ in a star cluster is related its total mass via $\mcl = \langle M \rangle N$, where $\langle M \rangle$ is the average mass of a particle (singles/binaries), which is defined by the mass distribution and the pairing properties of the binaries. For a cluster consisting of single stars only, the dynamical mass given by Eq.~(\ref{equation:spitzer}); the determination of $\mdyn$ is not dependent on the value of $N$. This is {\em not} the case if binaries are present. In this case, a larger $N$ (or larger total mass) results in an increased $\sigmapart$, while the orbital motion of the binaries (reflected in $\sigmaorb$) remains unaffected. Clusters with a large $N$ are thus less affected by binarity, and allow for a more accurate determination of $\mdyn$ than those with small $N$.

%_____________________________________________________________________________

\subsection{The half-mass radius $\halfmass$ and stellar density $\rho_{\rm hm}$}

 \begin{figure}[tb]
 \includegraphics[width=\columnwidth]{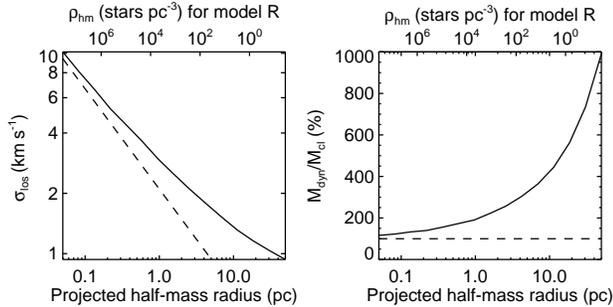}
 \caption{The effect of the half-mass radius $\halfmass$ and density $\rho_{\rm hm}$ on the measured line-of-sight velocity dispersion $\sigmalos$ and the overestimation of the dynamical mass of the cluster. The dashed lines indicate the results calculated from the centre-of-mass motion, and the solid curve the results for the {\em measured} line-of-sight velocities.  } %% no full stop at the end of caption
 \label{figure:density}
 \end{figure}

For a cluster with a given total mass and virial ratio, the velocity dispersion depends strongly on its size: $\sigmapart \propto \halfmass^{-1/2}$; see Eq.~(\ref{equation:spitzer}). Particles in a large (i.e., sparse) cluster move slower in the potential than in a tight, high-density cluster. The orbital motion of binary systems is independent of the cluster properties. In a low-density cluster the binary orbital motion is thus expected to dominate $\sigmalos$, while the presence of binaries is negligible in high-density clusters. The dependence on $\halfmass$ and the average density within the half-mass radius $\rho_{\rm hm}$ is illustrated in Fig.~\ref{figure:density}. Note that the dynamical mass of a distant massive OB~association ($\mcl \approx 10^4\msun$; $\halfmass \approx 20$~pc) can be overestimated by almost an order of magnitude, if its binaries are not properly taken into account.

%_____________________________________________________________________________

\subsection{The virial ratio $Q$}

After star formation the remaining gas from which the stars have formed is ejected by stellar winds and supernova explosions. During this phase a star cluster may loose a significant fraction of its mass, resulting in a reduced gravitational potential. The cluster is now out of virial equilibrium, possibly even unbound, and starts expanding.
Although Eq.~(\ref{equation:spitzer}) assumes that the cluster is in virial equilibrium, a correction for expanding or contracting clusters is easily made. Let $Q \equiv -E_K/E_P$ be the virial ratio, with $E_K$ the total kinetic energy and $E_P$ the total potential energy of the cluster. Clusters with $Q<0.5$, $Q=0.5$ and $Q>0.5$ are contracting, in virial equilibrium, and expanding, respectively. For any value of $Q$, the relation between the true mass and the dynamical mass of a star cluster is given by $\mcl = (2Q)^{-1}\mdyn$.
\cite{goodwinbastian2006} define the effective star forming efficiency (eSFE) $\epsilon$ as the star-forming efficiency that one would derive from the virial ratio under the assumption that the star-forming cloud was originally in virial equilibrium: $Q=(2\epsilon)^{-1}$. Under this assumption, the dynamical mass overestimation is given by $\mdyn/\mcl=\epsilon^{-1}$.

%_____________________________________________________________________________

 \begin{figure}[tb]
 \includegraphics[width=\columnwidth]{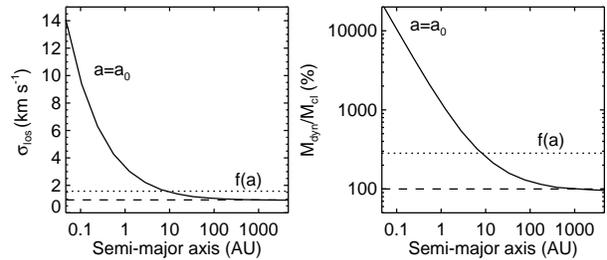}
 \caption{The effect of the binary semi-major axis $a$ on the measured line-of-sight velocity dispersion $\sigmalos$ and the overestimation of the dynamical mass of the cluster. All cluster properties are listed in Table~\ref{table:twomodels}. The solid curves show the results for clusters in which all binaries have an identical $a=a_0$, with $a_0$ along the horizontal axis. The dotted curve indicate the results for two realistic distributions: $f(a) \propto a^{-1}$ (\"{O}pik's law) and the \cite{duquennoy1991} period distribution. The results for both distributions are indistinguishable in this figure. The dashed lines indicate the results calculated from the centre-of-mass motion.} %% no full stop at the end of caption
 \label{figure:sma}
 \end{figure}

\section{Dependence on binary population properties} \label{section:binaryproperties}

The reliability of a dynamical mass determination depends not only on the properties of the cluster (which are reflected in $\sigmapart$), but also the properties of the binary population (which are reflected in $\sigmaorb$), in particular on the size of the binary orbits and the binary fraction. The effect of each of these binary population properties is discussed in the sections below.

%_____________________________________________________________________________

\subsection{The semi-major axis distribution $f(a)$}

The distribution over semi-major axes (or periods) is one of the most important parameters that affect the interpretation of the observed $\sigmalos$, as the orbital velocity of a component is a binary system is proportional to $a^{-1/2}$. Clusters containing wide binaries are less affected by binarity than those containing tight binaries. This effect is clearly shown in Fig.~\ref{figure:sma}. In binary-dominated clusters, which have $\sigmalos \approx \sigmaorb$, the velocity dispersion scales as $\sigmalos \propto a^{-1/2}$, and therefore $\mdyn/\mcl \propto a^{-1}$. In the particle-dominated case (which occurs if most binaries are wide), the dynamical mass is a good representation of the true cluster mass. The intermediate case occurs approximately at $\log (a/\halfmass) \approx -4.7$, where $\halfmass$ is the half-mass radius of the cluster. Note that the two most commonly used orbital size distributions, $f(a) \propto a^{-1}$ \citep[\"{O}pik's law; e.g.][]{poveda2004} and the log-normal period distribution \citep[][]{duquennoy1991} practically give the same results.

%_____________________________________________________________________________

\subsection{The binary fraction $\binfrac$}

 \begin{figure}[tb]
 \includegraphics[width=\columnwidth]{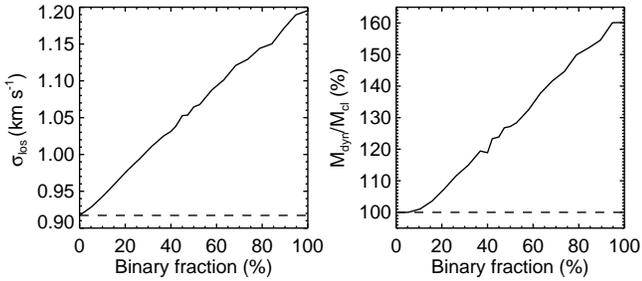}
 \caption{The effect of the binary fraction $\binfrac$ on the measured line-of-sight velocity dispersion $\sigmalos$ and the overestimation of the dynamical mass of the cluster. Model parameters are listed in Table~\ref{table:twomodels}. The dashed lines indicate the results calculated from the centre-of-mass motion, and the solid curve the results for the {\em measured} line-of-sight velocities.} %% no full stop at the end of caption
 \label{figure:bf}
 \end{figure}

The binary fraction is an important parameter, as it determines the relative weight that is given by $\sigmapart$ (the single stars) and $\sigmaorb$ (the binaries) to the measured value of $\sigmalos$. The dynamical mass of a star cluster with a low binary fraction is expected to be only barely overestimated. As the binary fraction increases, the overestimation gradually becomes larger. In the intermediate-case, the dynamical mass overestimation scales more or less linearly with $\binfrac$, which is illustrated in Fig.~\ref{figure:bf}.

%_____________________________________________________________________________

\subsection{The eccentricity distribution $f(e)$}

Stars in an eccentric orbit spend most of their time near apastron, where their velocity is small, and a small fraction of their time near periastron, where their velocity is large. The average velocity of a set of eccentric binaries at a certain point in time is therefore relatively small. For a cluster with highly eccentric orbits, the contribution of $\sigmaorb$ to $\sigmalos$ is therefore smaller than for a similar cluster with circular binaries. The dynamical mass overestimation thus decreases with increasing average eccentricity. However, the effect of an incorrectly adopted eccentricity distribution on the calculated $\mdyn$ is small, as compared to that of the uncertainty in $f(a)$, $\binfrac$, and the selection effects. For model~R our simulations show $\mdyn/\mcl = 80\%$ for a cluster with circular binaries, while $\mdyn/\mcl = 50\%$ if all binaries have $e=0.95$. 

%_____________________________________________________________________________

\subsection{The mass ratio distribution $f(q)$}

Equal-mass stars in a binary system orbit each other at equal velocities. For a very low mass ratio binary, the massive primary star barely moves, while the low-mass companion orbits at high velocity. Depending on which star is measured, the orbital velocity contribution to $\sigmalos$ can be small or large. Most frequently the velocity of the brightest, and therefore most massive star is measured. A higher average mass ratio thus leads to a larger dynamical mass overestimation. However, in practice the effect of an uncertainty in the mass ratio distribution is much smaller than for an uncertainty in $f(a)$, $\binfrac$ and the selection effects.

%_____________________________________________________________________________

\section{Selection effects}%\label{s:?}

Selection effects play an important role for the interpretation of the measured $\sigmalos$, in particular (i) the projected radius from the cluster centre at which the measurement is performed and (ii) the stellar mass range that is included in the observations. The line-of-sight velocity dispersion of the centres-of-mass at a certain projected distance $\rho$ from the cluster centre is
\begin{equation} \label{equation:lagrangiansigma}
  \sigmapart^2(\rho) = \frac{3\pi}{64}\frac{G\mcl}{\halfmass} \left( 1 + \frac{\rho^2}{\halfmass^2} \right)^{-1/2}
\end{equation}
\citep{heggiehut}. An expression for the dynamical mass overestimation as a function of $\rho$ is obtained by substituting Eq.~(\ref{equation:lagrangiansigma}) into Eq.~(\ref{equation:spitzer}):
\begin{equation}
  \frac{\mdyn}{\mcl} \approx \sqrt{2} \ \left( 1 + \frac{\rho^2}{\halfmass^2} \right)^{-1/2} \,.
\end{equation}
For velocity dispersions measured in the cluster centre this results in a mass overestimation by $\sim 40\%$. Measurements at the half-mass radius provide the correct $\mdyn$, while measurements in the cluster outskirts result in an underestimation of the mass. A possibly more important selection effect is introduced by the large brightness difference between stars of different masses. In reality, the determination of $\sigmalos$ is dominated by the properties of stars in a certain mass range (more specifically, a certain brightness range). The measured $\sigmalos$ may therefore not be representative for the cluster as a whole. This may result in a further dynamical mass overestimation \citep[e.g.,][]{kouwenhoven_sigma}. A detailed analysis of the selection effects is necessary to properly take these selection effects into account.

 \begin{figure}[tb]
 \includegraphics[width=\columnwidth]{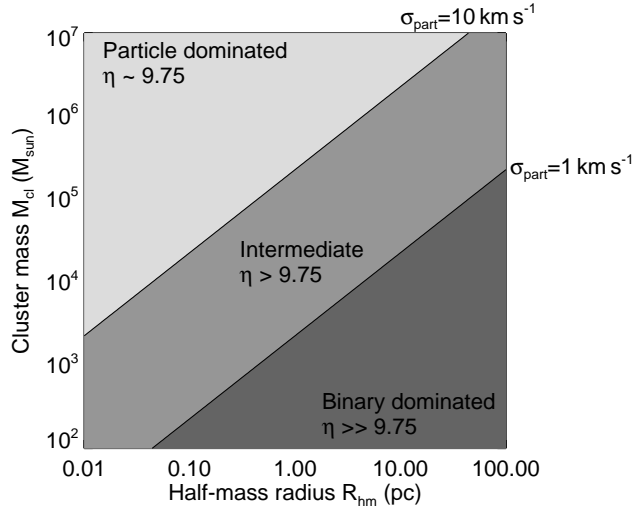}
 \caption{The effect of binarity on the dynamical mass determination for a cluster with a mass $\mcl$ and a half-mass radius $\halfmass$. For young massive clusters, binarity generally has a small effect on $\mdyn$, while for distant OB~associations the overestimation of $\mdyn$ can be up to an order of magnitude. } %% no full stop at the end of caption
 \label{figure:danger}
 \end{figure}

 \begin{figure}[tb]
 \includegraphics[width=\columnwidth]{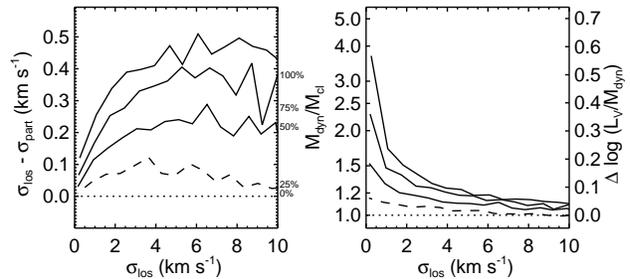}
 \caption{Given the {\em measured} line-of-sight velocity dispersion $\sigmalos$, what is the effect of binarity? Can the effect of binarity on the dynamical mass derivation be ignored? And if not, how severe is the overestimation of the inferred dynamical mass? For a given $\sigmalos$, the curves indicate the difference $\sigmalos-\sigmapart$ (left-hand panel), and the dynamical mass overestimation $\mdyn/\mcl$ (right-hand panel). Results are shown for models with $\binfrac=0\%$ (dotted curves), $\binfrac=25\%$ (dashed curves), and for $50\%$, $75\%$ and $100\%$ (solid curves). We additionally indicate the shift $\Delta \log (L_V/\mdyn)$ in the $(L_V/\mdyn)$ vs. age diagram that is introduced by the presence of binaries. } %% no full stop at the end of caption
 \label{figure:measured}
 \end{figure}

%_____________________________________________________________________________

\section{When can binaries be ignored?} \label{section:danger}

When studying a star cluster in detail, it is important to find out how reliable the measured dynamical mass is. Fig.~\ref{figure:danger} shows for clusters of different size $\halfmass$ and mass $\mcl$ whether they are particle-dominated, intermediate-case, or binary-dominated. The figure shows that the low-density OB~associations are generally binary-dominated; their dynamical mass as obtained from integrated spectral lines would significantly overestimated their true mass. Young massive clusters, with masses of $10^{5-6} \msun$ and typical half-mass radii of a few parsec, are of the intermediate case. Their dynamical masses are overestimated by a few per cent. From a practical point-of-view, Fig.~\ref{figure:measured} shows how the {\em measured} velocity dispersion $\sigmalos$ should be interpreted, and how the results depend on the intrinsic binary fraction. In the right-hand panel of Fig.~\ref{figure:measured} we additionally indicate the downwards shift $\Delta \log (L_V/\mdyn)=\log(\mdyn/\mcl)$ in the luminosity-to-mass vs. age diagram as a results of binarity.

%_____________________________________________________________________________

\section{Conclusions}%\label{s:?}

The total mass of a star cluster is often inferred from its line-of-sight velocity dispersion $\sigmalos$ and half-mass radius $\halfmass$, assuming virial equilibrium. The latter approach includes the assumption that no binaries are present. However, most stars are known to be in binary systems, and their orbital motions provide an additional contribution to the measured $\sigmalos$. The latter value is now no longer representative for the (centre-of-mass) motion of the stars and binaries in the cluster potential. This effect may result in a significant dynamical mass overestimation. Depending on the magnitude of the dynamical mass overestimation, we distinguish between three types of clusters: particle-dominated ($\sigmaorb \ll \sigmapart$), intermediate-case ($\sigmaorb \approx \sigmapart$), and binary-dominated clusters ($\sigmaorb \gg \sigmapart$).

The orbital velocity of binaries is independent of the cluster properties (size, mass, etc.). Whether or not binary motion affects $\sigmalos$ is thus depends on the cluster properties. For clusters with a high stellar density (i.e, large $\mcl$ or small $\halfmass$), $\mdyn$ is generally unaffected. In the latter case $\sigmapart > 10$~\kms{}. The dynamical mass overestimation increases strongly with (i) a higher binary fraction and (ii) a smaller average orbital size. The dependence on the mass ratio distribution and eccentricity distribution is small: $\Delta(\mdyn/\mcl) \la 5\%$.
We additionally show that observing a certain subset of the cluster introduces a selection effects, which may result in a further mass overestimation by up to 40\%. Analysis of the brightest (generally massive) stars in the cluster may further overestimate the dynamical mass. The dynamical mass overestimation can be negligible for the most massive clusters, while it may overestimate the true mass by up to an order of magnitude for sparse OB associations. The full results of this study have been presented in \cite{kouwenhoven_sigma}. A follow-up paper, which treats the selection effects properly, is in preparation.

%\subsection{}%\label{ss:?}
%\subsubsection{}%\label{sss:?}

%% Math 
%
%\begin{eqnarray}%\label{eqn:?}
%\\ \nonumber
%\end{eqnarray}
%
%\begin{equation}%\label{eqn:?}
%\end{equation}

%% Table (two-column)
%
% \begin{table*}%%[tb]
% \small
% \caption{Caption} %% no full stop at the end of caption
%  \label{tbl:?}
% \begin{tabular}{}
% \tableline  %% rule at top
% \tablenotemark{a} 
% <entries>
% \tableline %% rule at bottom
% \end{tabular}
%
%% Any table notes must follow the \end{tabular} command. 
% \tablenotetext{a}{}
% \tablecomments{}
% \end{table*}

%% Table (one-colum)
%
% \begin{table}
% \caption{} %% no full stop at the end of caption
% \label{tbl:?}
% \begin{tabular}{}
% \tableline  %% rule at top
% \tablenotemark{a} 
% <entries>
% \tableline %% rule at bottom
% \end{tabular}
% \end{table}

%% Deluxe tabel (refer to AASTeX documentation)
%
% \begin{deluxetable}{ccrrrrrrrrcrl}
% \tabletypesize{\scriptsize}
% \rotate
% \tablecaption{}% %% no full stop at the end of caption
% \label{tbl:?} 
% \tablewidth{0pt}
% \tablehead{\colhead{}}
% \startdata
% <entries>
% \enddata
%
%% Text for table notes should follow after the \enddata but before
%% the \end{deluxetable}. Make sure there is at least one \tablenotemark
%% in the table for each \tablenotetext.
% \tablecomments{}
% \tablenotetext{a}{}
% \tablenotetext{b}{}
% \end{deluxetable}

%% Figure 
%
% \begin{figure}%[tb]
% \includegraphics{}
% \includegraphics[width=\columnwidth]{}
% \caption{} %% no full stop at the end of caption
% \label{fig:?}
% \end{figure}

%% Acknowledgements
\acknowledgments
M.B.N. Kouwenhoven was supported by PPARC/STFC (grant PP/D002036/1).

%% References
%% Please cite all reference entries in the article text using \cite or
%% equivalent command. 

%%%  Using BibTeX  (Name-Year style)
%
% \bibliographystyle{spr-mp-nameyear-cnd}  %% BibTeX style
% \bibliography{<bib data>}                %% BibTeX data

%\bibliographystyle{spr-mp-nameyear-cnd}
%\bibliography{bibliography.bib}

%% Non-BibTeX  (Name-Year style)
%
% \begin{thebibliography}{}
% \bibitem[\protect\citeauthoryear{<author>}{<year>]{ref:?}
%    <ref. entry>
% \bibitem[\protect\citeauthoryear{<author>}{<year>]{ref:?}
%    <ref. entry>
% \end{thebibliography}

%\nocite{*}
%\bibliography{myref}
%\bibliography{biblio-u1}
%%%%\bibliographystyle{Spr-mp-nameyear-cnd}
%\bibliography{granada}
\bibliographystyle{aa}
\bibliography{granada}

\end{document}